\documentclass{PoS}

\usepackage{mathrsfs}
\usepackage{amssymb}
\usepackage{amsmath}
\usepackage{graphicx}
\usepackage{amsfonts,amsbsy}
\usepackage{color}
\usepackage{nicefrac}
\usepackage{xcolor}
\usepackage{comment}

\newcommand{\mbf}{\mathbf}

\newcommand{\mrm}{\mathrm}

\newcommand{\Tr}{\mrm{Tr}}
\newcommand{\HTL}{\mrm{HTL}}

\newcommand{\Q}{Q}

\newcommand{\fig}{Fig.~}

\newcommand{\se}{Sec.~}

\newcommand{\ud}{\mathrm{d}}

\title{Probing spectral properties of the QGP with real-time lattice simulations}

\ShortTitle{Probing spectral properties of the QGP with real-time lattice simulations}

\author{\speaker{Kirill Boguslavski}\thanks{The authors are grateful to J.\ Berges, A.\ Pi\~{n}eiro Orioli, T.\ Gasenzer, J.\ Ghiglieri, A.\ Mazeliauskas, J.\ Pawlowski, A.\ Rebhan, C.~A.\ Salgado and M.\ Strickland for valuable discussions and would like to thank Asier Pi\~{n}eiro Orioli for sharing with us a method to improve the efficiency of our simulations in a private communication. T.~L.\ is supported by the Academy of Finland, projects No. 267321 and No. 303756. This work is supported  by the European Research Council, grant ERC-2015-CoG-681707. J.~P.\ is supported by the Jenny and Antti Wihuri Foundation. J.~P.\ acknowledges support for travel from Magnus Ehrnrooth foundation. K.~B.\ and J.~P.\ would like to thank CERN and its Theory group for hospitality during part of this work. The authors wish to acknowledge CSC - IT Center for Science, Finland, for computational resources. }\\
        Department of Physics, P.O.~Box 35, 40014 University of Jyv\"{a}skyl\"{a}, Finland\\
        E-mail: \email{kirill.boguslavski@jyu.fi}}

\author{Aleksi Kurkela\\
        Theoretical Physics Department, CERN, Geneva, Switzerland\\
        Faculty of Science and Technology, University of Stavanger, 4036 Stavanger, Norway\\
        E-mail: \email{aleksi.kurkela@cern.ch}}
        
\author{Tuomas Lappi\\
        Department of Physics, P.O.~Box 35, 40014 University of Jyv\"{a}skyl\"{a}, Finland\\
        Helsinki Institute of Physics, P.O.~Box 64, 00014 University of Helsinki, Finland\\
        E-mail: \email{tuomas.v.v.lappi@jyu.fi}}
        
\author{Jarkko Peuron\\
        European Centre for Theoretical Studies in Nuclear Physics and Related Areas (ECT*) and Fondazione Bruno Kessler, Strada delle Tabarelle 286, I-38123 Villazzano (TN), Italy \\
        E-mail: \email{jpeuron@ectstar.eu}}

\abstract{We present a new method to obtain spectral properties of a non-Abelian gauge theory in the region where occupation numbers are high. The method to measure the (single-particle) spectral function is based on linear response theory and classical-statistical lattice simulations. Although we apply it to a system far from equilibrium in a self-similar regime, the extracted spectral function can be understood within the hard thermal loop (HTL) formalism and can thus be connected to thermal equilibrium at high temperatures. This allows us to obtain quantities like the lifetime of quasiparticles that are beyond the leading order and difficult to compute within HTL. The approach has the potential to measure transport coefficients, to study the earliest stages of heavy-ion collisions in a controlled way and it can be employed beyond the range of validity of HTL.}

\FullConference{International Conference on Hard and Electromagnetic Probes of High-Energy Nuclear Collisions\\
		30 September - 5 October 2018\\
		Aix-Les-Bains, Savoie, France}

\begin{document}

\section{Introduction}

The knowledge of the spectral function $\rho(t,\omega,p)$ is crucial to understand the underlying excitation spectrum of a system. In \cite{Boguslavski:2018beu}, we developed a method to obtain spectral functions and other unequal-time correlation functions non-perturbatively in non-Abelian gauge theories out of equilibrium, which uses the formalism of \cite{Kurkela:2016mhu} and is presented in this Proceeding. 

The method can be applied to highly occupied systems or when Yang Mills fields are large. Such situations occur at initial stages of ultra-relativistic heavy-ion collisions (URHICs) for weak couplings. After the collision, the out-of-equilibrium Quark-Gluon plasma (QGP) can be described by large classical Yang Mills fields that transform into a highly occupied plasma via instabilities \cite{ClassGlasmaInstab}. After that, its pre-hydrodynamic evolution can be described by effective kinetic theory \cite{BottomUp}. 
Some phenomenologically relevant questions of the initial stages can be addressed with the new method. These involve the instability phase and the transition to the highly occupied plasma, the details of which are not fully understood yet. 
A detailed knowledge of unequal-time correlation functions at initial stages is also necessary to study the early-time evolution of the jet quenching parameter $\hat{q}$, the heavy-quark diffusion coefficient $\kappa$ and of transport coefficients. 

In this work, we will explain the main working principle of the method and we will apply it to an isotropic self-similar non-Abelian plasma. Because of a resulting scale separation of soft and hard modes, the hard thermal loop (HTL) formalism is applicable \cite{HTL} and we will compare our numerical results to it. The comparison has a two-fold goal. On the one hand, we will check to what extent leading order (LO) HTL expressions describe the far-from-equilibrium system. On the other hand, interpreting our results as a non-perturbative simulation of HTL, we can compute quantities like the quasi-particle damping rate in a controlled way that are difficult to compute within HTL. This provides a deep insight into thermal equilibrium for high temperatures $T \gg \Lambda_{\rm{QCD}}$.

\section{New method to extract spectral functions}
\label{sec_method}

The far-from-equilibrium evolution of highly occupied non-Abelian plasmas of an $\mrm{SU}(N_c)$ gauge theory can be efficiently described using classical-statistical (CSA) lattice simulations. Classical gauge and chromo-electric fields $A_j(t,\mbf x)$ and $E^j(t,\mbf x)$ with Lorentz index $j = 1,2,3$ are discretized on lattices $N_s^3$ with lattice spacing $a_s$. Their evolution is determined by classical Yang-Mills equations of motion (EOM) in temporal gauge $A_0 = E^0 = 0$.\footnote{More precisely, gauge fields are replaced by the gauge covariant link fields $U_j(t,\mbf x) \approx \exp(i g a_s A_j(t,\mbf x))$ to render the evolution equations also gauge co-variant. See, e.g., \cite{Boguslavski:2018beu,SelfSimIso} for more details.}

\begin{figure}[t]
	\centering
	\includegraphics[scale=0.27]{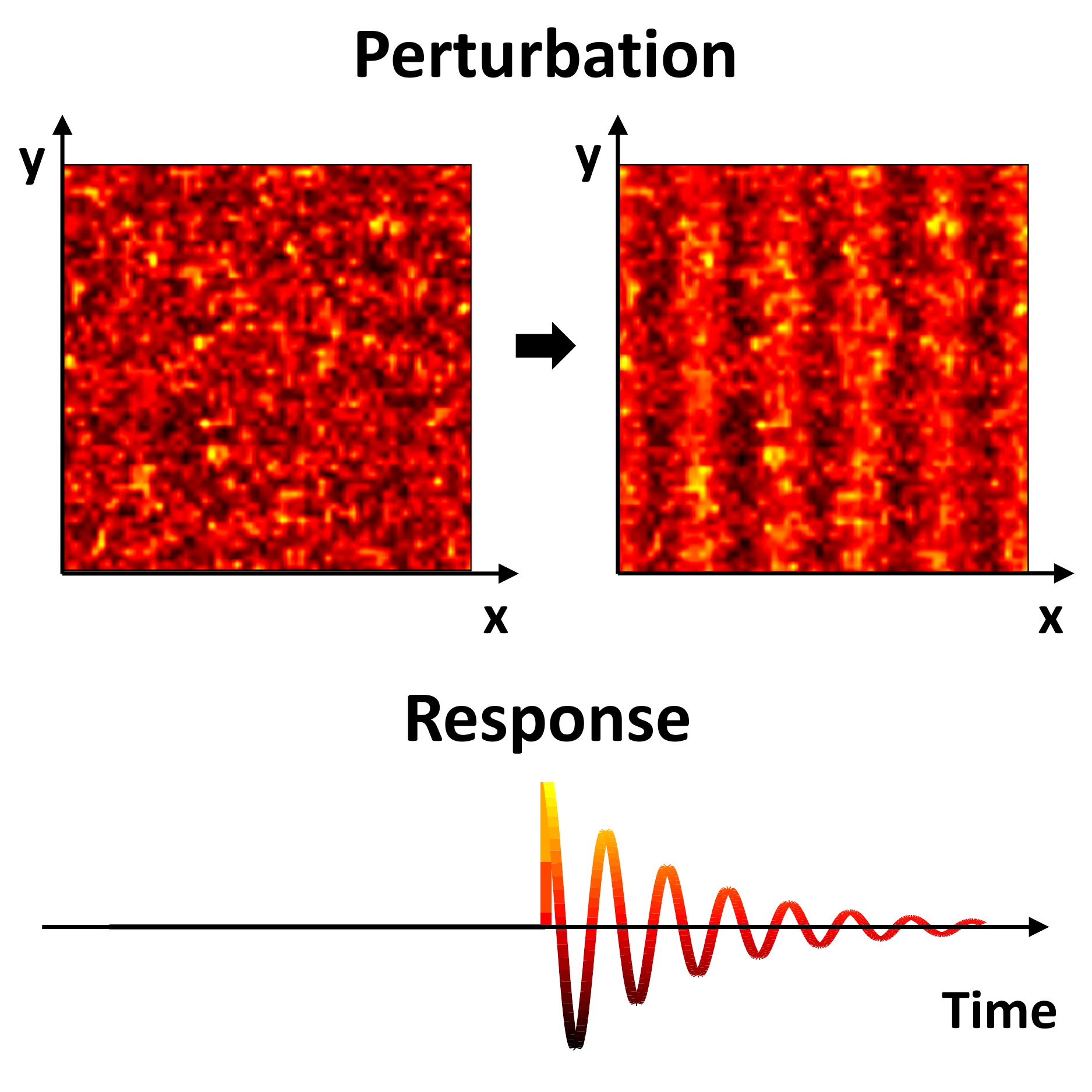} \quad
	\includegraphics[scale=0.27]{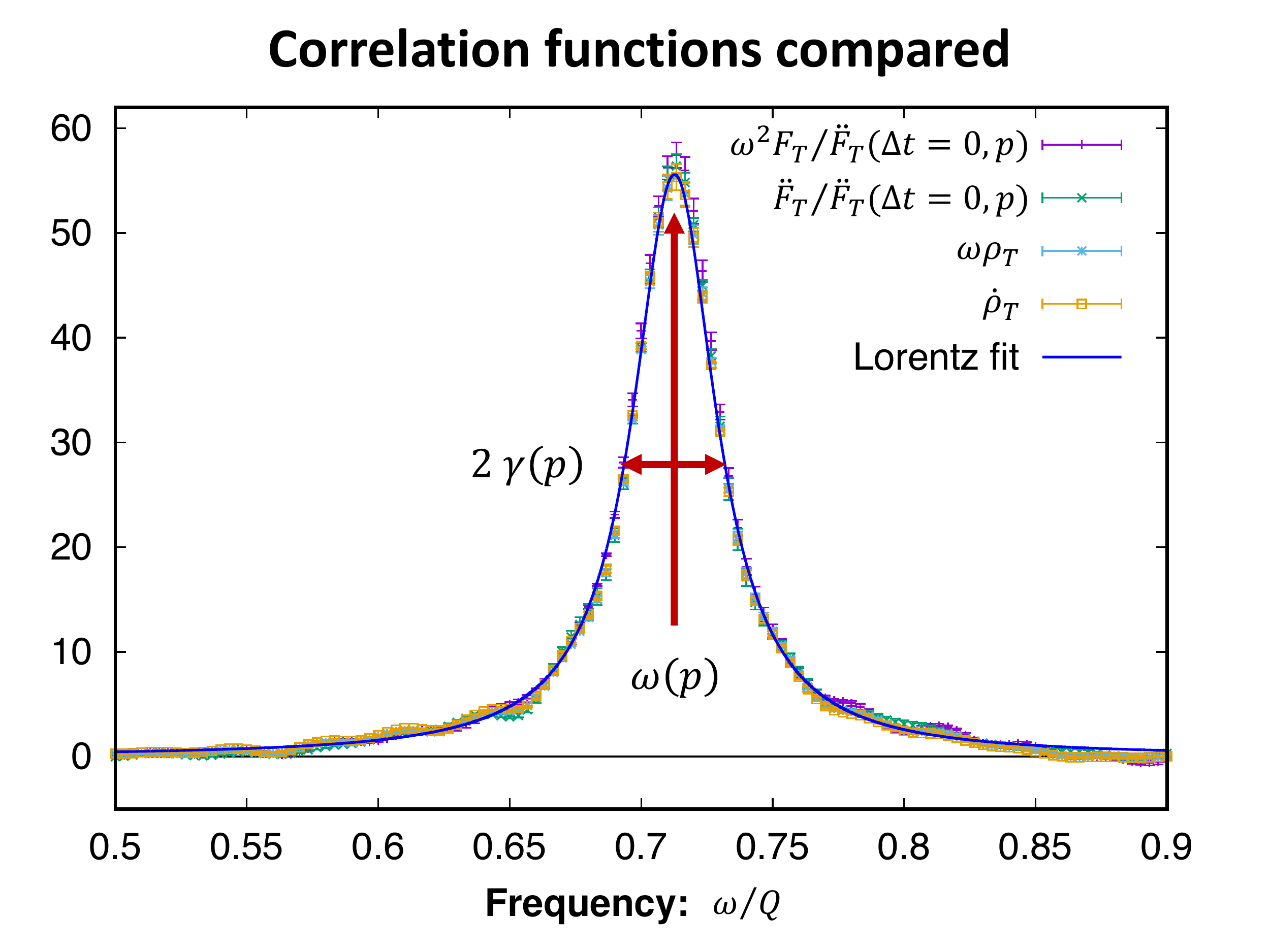} 
	\caption{{\bf Left:} Illustration of the new method to extract spectral functions on a real-time lattice, as described in \se\ref{sec_method}. {\bf Right:} A quasiparticle excitation shown for $p = 0.7\, \Q$ for different correlation functions. A fit to $a(t,p) / ((\omega - \omega(t,p))^2 + \gamma^2(t,p))$ is included. (figures developed for the highlighted article \cite{Boguslavski:2018beu})}
	\label{fig_spectral_principle}
\end{figure}

A new technique to measure spectral functions of highly occupied non-Abelian plasmas has been recently developed in \cite{Boguslavski:2018beu}. It extends the CSA method by employing linear response theory with linearized EOM from \cite{Kurkela:2016mhu}. We will only sketch the working principle here, that is illustrated in the left panel of \fig\ref{fig_spectral_principle}. The upper parts of the panel show snapshots of $\sum_j \Tr(E^j E^j)$ at $z=0$ at a time $t'$ before and after a perturbation with the instantaneous source $j \propto e^{i\, \mbf p\, \mbf x}\,\delta(t - t')$ with momentum $\mbf p$, chosen in the $x$ direction. For the subsequent evolution, the fields are split into a part not affected by the perturbation and the response field caused by the perturbation, i.e., $E^j(t,\mbf x) \mapsto E^j(t,\mbf x) + e^j(t,\mbf x)$ and similar for $A_j$. While the unperturbed fields $E_j(t,\mbf x)$ follow conventional Yang-Mills EOM, the evolution of the response fields $e^j$ and $a_j$ at times $t > t'$ is determined by linearised EOM \cite{Kurkela:2016mhu}. A typical evolution of the response $\langle e^j \rangle$ is shown in the lower part of the right panel of \fig\ref{fig_spectral_principle}. Using linear response theory, these oscillations can be connected to the retarded propagator $G_{R}$ and the spectral function $\rho$ in Fourier space
\begin{align}
 \label{eq_linresp_GR}
 \langle a_j(t,\mbf p) \rangle = G_{R,jk}(t,t',\mbf p)\,j^k(t',\mbf p)\;, \qquad G_{R,jk}(t,t',\mbf p) = \theta(t-t')\,\rho_{jk}(t,t',\mbf p)\;,
\end{align}
where $j^k(t',\mbf p)$ is the remaining part of the source without the time Delta function. The spectral function can be efficiently extracted from \eqref{eq_linresp_GR} by use of statistical methods that have been developed in \cite{PineiroOrioli:2018hst,Boguslavski:2018beu} and allow to perturb the system at several momentum modes simultaneously. Moreover, transverse and longitudinal polarizations $\rho_T$ and $\rho_L$ can be distinguished.

\section{First application and numerical results}

When a bosonic theory is highly occupied $f \sim 1/g^2$ for typical modes at weak couplings $g^2 \ll 1$, where $f(t,p)$ is the (quasi-particle) distribution function, it often approaches a non-thermal fixed point (NTFP) during its out-of-equilibrium evolution \cite{NTFP}. In the proximity of a NTFP, the evolution becomes insensitive to details of the initial conditions and slows down considerably following a self-similar form $f(t,p) = (\Q t)^\alpha f_s((\Q t)^\beta p)$ with a constant scale $\Q$ and universal scaling exponents $\alpha$, $\beta$ and scaling function $f_s(p)$.
As a first application of our method, we choose a NTFP emerging in isotropic $\mrm{SU}(N_c)$ gauge theory in $(3+1)$ dimensions. It has the universal scaling exponents $\beta = -1/7$, $\alpha = 4\beta$ and involves a power law $f(p) \sim (p/\Q)^{-\kappa}$ with $\kappa \simeq 4/3 \sim 1$ for momenta below a hard scale $p \lesssim \Lambda(t)$ \cite{SelfSimIso}. The hard scale 
grows with time while the mass decreases.\footnote{The mass can be computed within the HTL framework as $m^2(t) \approx 2 N_c \int \ud^3 p/(2\pi)^3 g^2 f(t,p)/p$.} 
This leads to a scale separation that grows with time as $\Lambda(t) / m(t) \sim (\Q t)^{2/7}$. Because of this, the HTL framework \cite{HTL} becomes applicable, which organizes perturbation theory in terms of $m / \Lambda \ll 1$. Our results will be therefore compared to the corresponding HTL expressions. 

Using the new method at the self-similar system, we show a typical spectral function $\partial_t \rho_T = \dot{\rho}_T$ in the right panel of \fig\ref{fig_spectral_principle} as a function of frequency $\omega$ at fixed momentum. One finds a quasi-particle peak that follows the Lorentz function $\propto ((\omega - \omega(t,p))^2 + \gamma^2(t,p))^{-1}$. 
The position of the maximum $\omega(t,p)$ corresponds to the dispersion relation (the quasi-particles' energy) and the peak width $\gamma(t,p)$ to the damping rate (inverse lifetime). Apart from the spectral function, we also show the statistical correlation function $F_{jk}(t,\Delta t,p) = \langle A_j(t,\mbf p) A_j^*(t',\mbf p) \rangle/((a_s N_s)^3 (N_c^2-1))$ and similarly $\ddot{F}^{jk}$ defined using $E^j(t,\mbf p)$. While in thermal equilibrium, it is equivalent to the spectral function because of the fluctuation-dissipation relation $\ddot{F} = T\, \dot{\rho}$ for small $\omega, p \ll T$ below the temperature, we find $\omega^2 F \approx \ddot{F}$, $\omega \rho \approx \dot{\rho}$ and a generalized fluctuation-dissipation relation far from equilibrium $\ddot{F}_{T}(t,\omega,p) \approx \ddot{F}_{T}(t,\Delta t = 0,p)\; \dot{\rho}_{T}(t,\omega,p)$, and similar for the longitudinal case.

\begin{figure}[t]
	\centering
	\includegraphics[scale=0.42]{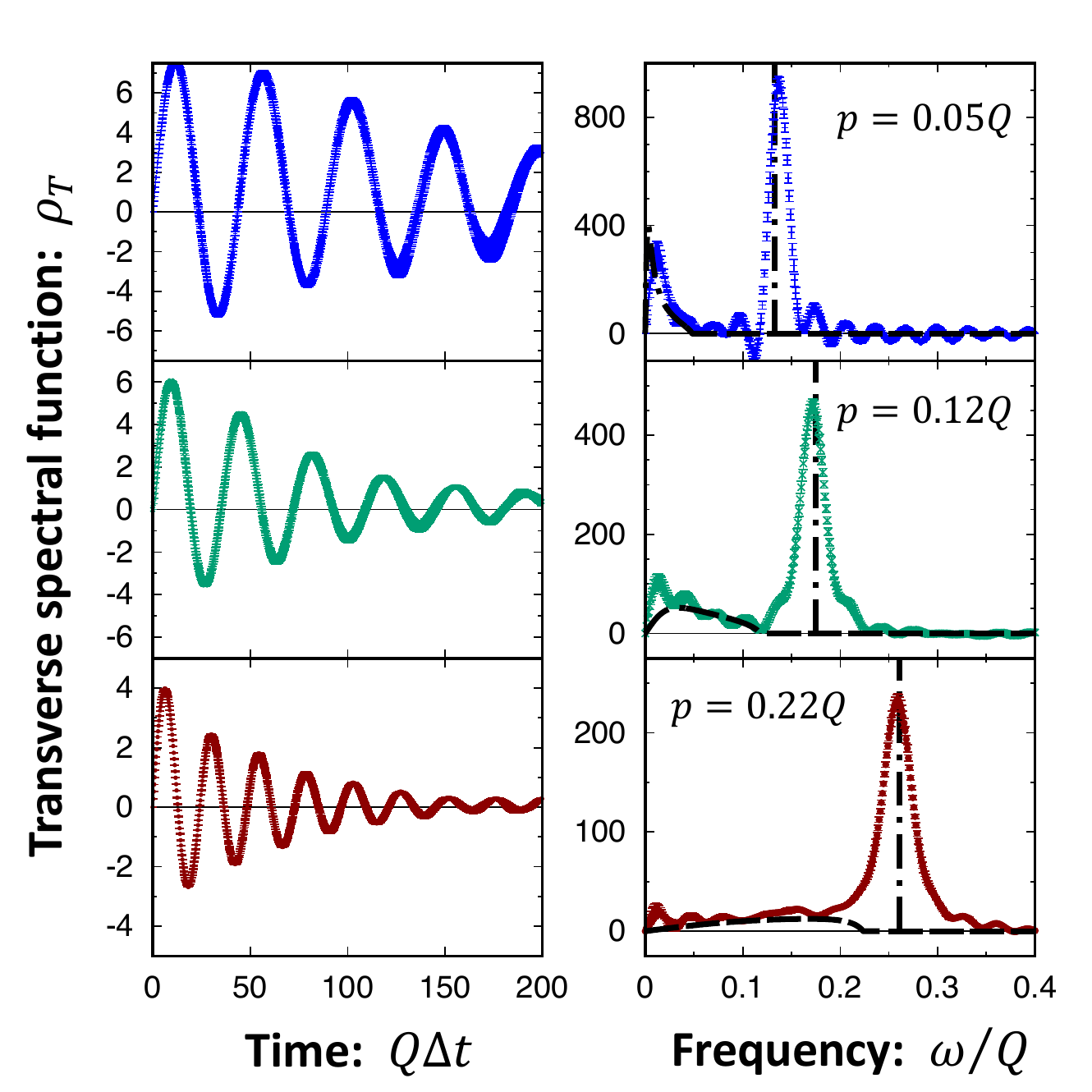} $\qquad$
	\includegraphics[scale=0.5]{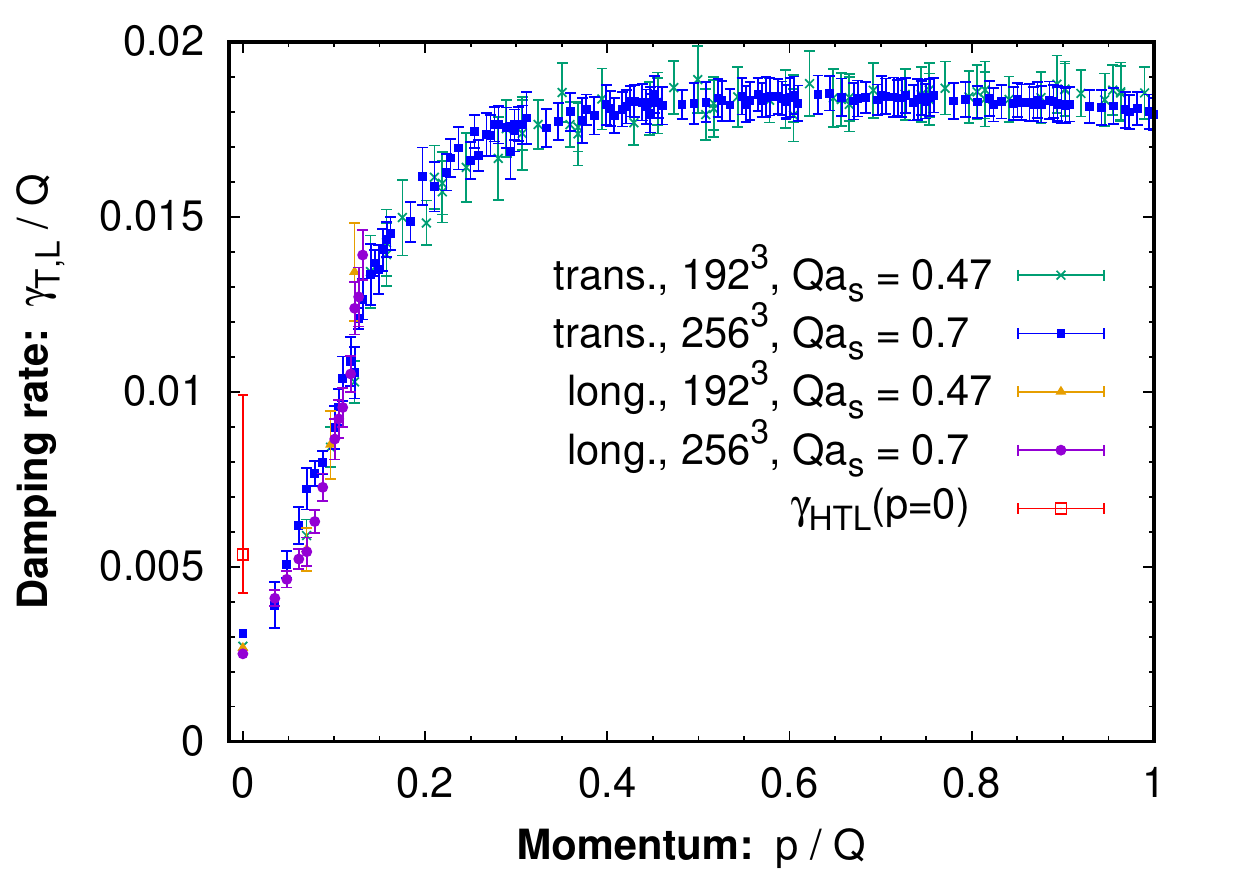}
	\caption{{\bf Left:} Spectral function $\rho_T$ as a function of $\Delta t$ and $\omega$. Dashed lines are computed with HTL theory at LO. 
	{\bf Right:} Damping rates $\gamma_{T,L}$ of the quasi-particles in $\rho_{T,L}$ as functions of momentum for different discretizations. The point $\gamma_{\HTL}(p=0)$ is computed using HTL \cite{HTL}. Although consistent with the curves, $\gamma_{\HTL}(p\rightarrow \infty)$ is omitted because of huge error bars due to the magnetic scale. (both figures taken from \cite{Boguslavski:2018beu})}
	\label{fig_spectral}
\end{figure}

Also more complex structures are encoded in $\rho(t,\omega,p)$. This is shown in the left panel of \fig\ref{fig_spectral} for the transversely polarized spectral function $\rho_T$ at different momenta and fixed time $t$. It is shown as a function of the time difference $\Delta t = t-t'$ (with $\Delta t \ll t$) and frequency $\omega$ from Fourier transforming with respect to $\Delta t$. In addition to the quasi-particle peak, one finds an enhancement for $|\omega| \leq p$ corresponding to the Landau damping of the system. Interestingly, Landau damping and the peak position (dispersion relation) agree well with the curves computed in the HTL formalism at LO. 

Their main difference lies in the peak width, which is neglected in the LO HTL curves since it is a quantity at next-to-leading order (NLO) in that formalism. The extracted finite width, i.e., the damping rates are shown in the right panel of \fig\ref{fig_spectral} for both polarizations as functions of momentum. One observes that they increase at $p \lesssim 0.15\,\Q \approx m$ while flattening at higher momenta. Remarkably, they coincide at low momenta irrespective of the polarization. Moreover, the value at $p=0$ is roughly consistent with the HTL value computed at NLO \cite{HTL}.
This is the first time that the damping rates were determined for a wide momentum range. Extending the known HTL results, this finding improves our knowledge of the high-temperature thermal state. It also shows that the system is dominated by long-lived quasi-particles, enabling a kinetic description.

\section{Conclusion}
\label{sec_conclusion}

We have presented here our new technique developed in \cite{Boguslavski:2018beu,Kurkela:2016mhu} where linear response theory is combined with classical simulations to obtain the spectral function $\rho$ of highly occupied non-Abelian systems. Applying it to the self-similar regime of an isotropic non-Abelian plasma, we found that $\rho$ involves a Landau damping region and a quasi-particle peak and is connected to the symmetric statistical propagator $F$ in terms of a generalized fluctuation-dissipation relation. Our results for $\rho$ are mostly consistent with HTL computations at LO. The main difference is that we were able to extract the full momentum dependence of the quasi-particle width $\gamma(p)$ for the first time, extending HTL results for $p=0$ computed at NLO. 

This is a promising method to address the questions discussed in the introduction. For heavy-ion collisions, it can be used to study the earliest stages of the evolution by distinguishing between strong classical fields and their unstable modes numerically. Studying unequal-time correlation functions such as $F$ and $\rho$, also transport properties can be extracted at initial stages. Moreover, extending this method to other theories, it becomes a useful tool to understand the underlying microscopic dynamics of systems far from equilibrium.


\begin{thebibliography}{99}

\bibitem{Boguslavski:2018beu} 
  K.~Boguslavski, A.~Kurkela, T.~Lappi, J.~Peuron,
  Phys.\ Rev.\ D {\bf 98}, 014006 (2018).

\bibitem{Kurkela:2016mhu} 
  A.~Kurkela, T.~Lappi, J.~Peuron,
  Eur.\ Phys.\ J.\ C {\bf 76}, 688 (2016).
  
\bibitem{ClassGlasmaInstab}
  A.~Krasnitz, R.~Venugopalan, 
  Nucl.\ Phys.\ B {\bf 557}, 237 (1999); 
  Phys.\ Rev.\ Lett.\  {\bf 84}, 4309 (2000); 
  A.~Krasnitz, Y.~Nara, R.~Venugopalan, 
  Phys.\ Rev.\ Lett.\  {\bf 87}, 192302 (2001); 
  T.~Lappi, 
  Phys.\ Rev.\ C {\bf 67}, 054903 (2003); 
  S.~Mrowczynski, 
  Phys.\ Lett.\ B {\bf 314}, 118 (1993); 
  A.~Rebhan, P.~Romatschke, M.~Strickland, 
  Phys.\ Rev.\ Lett.\  {\bf 94}, 102303 (2005); 
  P.~Romatschke, R.~Venugopalan, 
  Phys.\ Rev.\ Lett.\  {\bf 96}, 062302 (2006); 
  J.~Berges, K.~Boguslavski, S.~Schlichting, 
  Phys.\ Rev.\ D {\bf 85}, 076005 (2012); 
  M.~Attems, A.~Rebhan, M.~Strickland, 
  Phys.\ Rev.\ D {\bf 87}, 025010 (2013); 
  J.~Berges, S.~Schlichting, 
  Phys.\ Rev.\ D {\bf 87}, 014026 (2013).
  
\bibitem{BottomUp}
  R.~Baier, A.~H.~Mueller, D.~Schiff, D.~T.~Son, 
  Phys.\ Lett.\ B {\bf 502}, 51 (2001); 
  J.~Berges, K.~Boguslavski, S.~Schlichting, R.~Venugopalan, 
  Phys.\ Rev.\ D {\bf 89}, 074011 (2014); 
  Phys.\ Rev.\ D {\bf 89}, 114007 (2014); 
  Phys.\ Rev.\ D {\bf 92}, 096006 (2015); 
  P.~B.~Arnold, G.~D.~Moore, L.~G.~Yaffe, 
  JHEP {\bf 0301}, 030 (2003); 
  A.~Kurkela, E.~Lu, 
  Phys.\ Rev.\ Lett.\  {\bf 113}, 182301 (2014); 
  A.~Kurkela, Y.~Zhu, 
  Phys.\ Rev.\ Lett.\  {\bf 115}, 182301 (2015); 
  A.~Kurkela, A.~Mazeliauskas, J.~F.~Paquet, S.~Schlichting, D.~Teaney, 
  arXiv:1805.00961 [hep-ph]; 
  arXiv:1805.01604 [hep-ph].

\bibitem{HTL}
  E.~Braaten, R.~D.~Pisarski, 
  Nucl.\ Phys.\ {\bf B} 337, 569 (1990); 
  Phys.\ Rev.\ {\bf D} 42, 2156 (1990); 
  R.~D.~Pisarski, 
  Phys.\ Rev.\ D {\bf 47}, 5589 (1993); 
  J.-P.\ Blaizot, E.~Iancu, 
  Phys.\ Rept.\ 359, 355 (2002).
  
\bibitem{SelfSimIso}
  J.~Berges, S.~Scheffler, D.~Sexty, 
  Phys.\ Lett.\ {\bf B} 681, 362 (2009); 
  A.~Kurkela, G.~D.~Moore, 
  JHEP {\bf 12}, 044 (2011); 
  Phys.\ Rev.\ {\bf D} 86, 056008 (2012); 
  S.~Schlichting, 
  Phys.\ Rev.\ {\bf D} 86, 065008 (2012); 
  M.~C.~Abraao York, A.~Kurkela, E.~Lu, G.~D.~Moore, 
  Phys.\ Rev.\ {\bf D} 89, 074036 (2014); 
  T.~Lappi, J.~Peuron, 
  Phys.\ Rev.\ D {\bf 95}, 014025 (2017).  
  
\bibitem{PineiroOrioli:2018hst}
  A.~Pi\~{n}eiro Orioli, J.~Berges, 
  arXiv:1810.12392 [cond-mat.quant-gas].
  
\bibitem{NTFP}
  J.~Berges, A.~Rothkopf, J.~Schmidt, 
  Phys.\ Rev.\ Lett.\  {\bf 101}, 041603 (2008); 
  J.~Berges, K.~Boguslavski, S.~Schlichting, R.~Venugopalan, 
  Phys.\ Rev.\ Lett.\  {\bf 114}, 061601 (2015); 
  A.~Pi\~{n}eiro Orioli, K.\ Boguslavski, J.\ Berges, 
  Phys.\ Rev.\ {\bf D} 92, 025041 (2015); 
  M.~Pr\"{u}fer {\it et al.}, 
  Nature {\bf 563}, 217 (2018); 
  S.~Erne, R.~B\"{u}cker, T.~Gasenzer, J.~Berges, J.~Schmiedmayer, 
  Nature {\bf 563}, 225 (2018).
  


\end{thebibliography}
\end{document}